\documentclass[preprint]{vgtc}               

\graphicspath{{figures/}{pictures/}{images/}{./}} 

\usepackage{times}                     

\usepackage{tabu}                      
\usepackage{booktabs}                  
\usepackage{lipsum}                    
\usepackage{mwe}                       

\usepackage{mathptmx}                  
\usepackage{tcolorbox}
\usepackage{xcolor}
\usepackage{listings}
\usepackage[table]{xcolor}
\usepackage{amsmath,amsfonts}
\usepackage{graphicx}  
\usepackage{array, multirow, booktabs}
\usepackage{wrapfig}
\usepackage[normalem]{ulem}
\usepackage[nocompress, sort]{cite}

\definecolor{design_goal_textbox}{HTML}{E2E2E2}

\definecolor{object_red}{HTML}{FF697D}
\definecolor{object_blue}{HTML}{66A5E1}
\definecolor{object_orange}{HTML}{FF9E46}

\definecolor{plot_red}{HTML}{FFA7B1}
\definecolor{plot_orange}{HTML}{FFD3AC}
\definecolor{plot_blue}{HTML}{9DC7F0}

\definecolor{lightred}{RGB}{255, 175, 177}
\definecolor{lightorange}{RGB}{255, 215, 174}
\definecolor{lightyellow}{RGB}{255, 250, 194}

\definecolor{kim_green}{HTML}{4CBB17}
\definecolor{ari_red}{HTML}{EC1313}

\definecolor{pending_user_study_res}{HTML}{2171DF}

\newcommand{\para}[1]{\vspace{0.35em}\noindent\normalsize\textbf{#1.}\xspace}
\newcommand{\parac}[1]{\vspace{0.35em}\noindent\normalsize\textbf{#1:}\xspace}

\newcommand{\dc}[2][1em]{%
  \raisebox{-0.5\dp\strutbox}{\includegraphics[height=#1]{#2}}%
}

\newcommand{\sys}[0]{Speech-to-Spatial\xspace}
\newcommand{\gpt}[0]{GPT-4.1\xspace}
\newcommand{\gemini}[0]{Gemini 2.5-Flash\xspace}

\newcommand{\suppl}[0]{Suppl. Material\xspace}

\newcommand{\audioc}[0]{\textit{Audio}\xspace}
\newcommand{\fullc}[0]{\textit{Full}\xspace}
\newcommand{\summaryc}[0]{\textit{Summary}\xspace}

\makeatletter
\newcommand{\recaptitle}[1]{%
  \noindent{\reset@font\sffamily\normalsize\vgtc@sectionfont\itshape #1.}%
}
\makeatother



\onlineid{1403}

\vgtccategory{System}

\vgtcinsertpkg

\usepackage{orcidlink}

\preprinttext{%
  \ifodd\value{page}%
    \parbox{0.9\textwidth}{%
      © 2026 IEEE. This is the author's version of the article that will appear at the IEEE Conference on Virtual Reality and 3D User Interfaces (IEEE VR). The final version of this record is available at: \href{https://doi.org/10.1109/VR67842.2026.00045}{\textcolor{blue}{\textit{10.1109/VR67842.2026.00045}}}
    }%
  \fi
}

\title{From Speech-to-Spatial: Grounding Utterances on A Live Shared View with Augmented Reality}

\author{Yoonsang Kim\thanks{e-mail:yoonsakim@cs.stonybrook.edu} %
\and Divyansh Pradhan\thanks{e-mail:divyansh.pradhan@stonybrook.edu} %
\and Devshree Jadeja\thanks{e-mail:devshreehardik.jadeja@stonybrook.edu} %
\and Arie E. Kaufman\thanks{e-mail:ari@cs.stonybrook.edu}}
\affiliation{\scriptsize Center for Visual Computing, Stony Brook University}

\teaser{
  \centering
  \includegraphics[width=\linewidth]{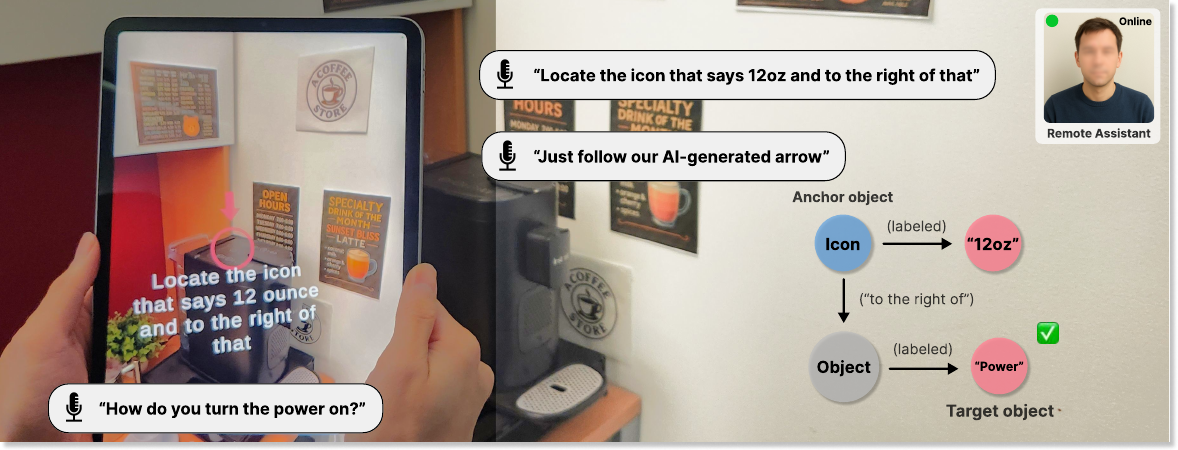}
  \vspace{-6mm}
  \caption{Concept illustration of \sys, disambiguating verbal descriptions of a referent and situating AR visual guiders leveraging object-centric graph, and LLM-based reasoning in a remote assistance scenario.}
  \label{fig:teaser}
}

\abstract{
We introduce \sys, a referent disambiguation framework that converts verbal remote-assistance instructions into spatially grounded AR guidance. Unlike prior systems that rely on additional cues (e.g., gesture, gaze) or manual expert annotations, \sys infers the intended target solely from spoken references (speech input). Motivated by our formative study of speech referencing patterns, we characterize recurring ways people specify targets (Direct Attribute, Relational, Remembrance, and Chained) and ground them to our object-centric relational graph. Given an utterance, referent cues are parsed and rendered as persistent in-situ AR visual guidance, reducing iterative micro-guidance (``\textit{a bit more to the right}'', ``\textit{now, stop.}'') during remote guidance. We demonstrate the use cases of our system with remote guided assistance and intent disambiguation scenarios. Our evaluation shows that \sys improves task efficiency, reduces cognitive load, and enhances usability compared to a conventional voice-only baseline, transforming disembodied verbal instruction into visually explainable, actionable guidance on a live shared view.
} 


\begin{document}


\firstsection{Introduction}

\maketitle
Remote assistance often relies on spoken instructions~\cite{druta2021review}, which are ambiguous when designating referents. Prior works show that language alone is insufficient, and phrases such as ``this one'' or ``over there'' remain under-specified unless paired with embodied cues such as gesture or gaze~\cite{bolt1980put, lee2024gazepointar, bai2020user, hu2025gesprompt}. Recent studies demonstrate that combining verbal and visual cues, aided by LLMs, enhances the clarity of verbal descriptions~\cite{jadon2024augmented, stover2024taggar, lee2025walkie}. Beyond gesture, relational context also grounds meaning: ontology-driven approaches model multi-dimensional attributes such as space, time, activity, actor, and object~\cite{kim2017ontology}, and relational priors and scene graphs yield dense, spatially consistent reconstructions~\cite{murai2025mast3r}. These works suggest that ambiguity can be reduced through structural analysis of how entities relate both spatially and semantically.

While graphs and multimodal cues clarify communication, eXtended Reality (XR) collaboration introduces practical challenges. Visualizing embodied gestures or remote users' avatars requires specialized hardware such as motion tracking and immersive Head-Mounted Displays (HMDs). These requirements slow the adoption of immersive solutions in remote technical support, where practitioners often fall back to 2D displays, hand-held device-based communication, or manual cursor annotations. Other domains employ chatbot-style AI with no direct human-to-human assistance or embodied presence. These constraints underline the need for lightweight approaches that can operate in speech-only conditions.

We propose \textbf{\sys} to address these challenges. \sys is a framework that grounds verbal-only instructions into Augmented Reality (AR) visual guidance (\cref{fig:teaser}). Building on the existing literature on the ambiguity of spatial expressions and our formative study, we categorize four recurring referential types -- Direct Feature-based, Relational, Memory-based, and Chained. \sys reconstructs a relational graph, maps spoken references, and auto-generates AR indicators that disambiguate spatial descriptions. This transforms transient utterances into persistent, spatially contextualized guidance--without requiring additional gear, cues, or manual human annotations.

We showcase the use cases of \sys across three collaborative scenarios: Remote maintenance instructions, Indoor navigation, and AI personal assistance. Our evaluation shows measurable gains in task efficiency, accuracy, reduced cognitive load, and heightened usability for \sys, compared to the voice-only assistance baseline. Our contributions are as follows:
\vspace{-1.5mm}
\begin{itemize}
\item \textbf{An end-to-end pipeline for speech disambiguation} We elucidate the intent behind speech-only instructions using graph-traversal reasoning and provide situated visual guidance through an AR indicator.
\vspace{-1mm}
\item \textbf{Spatial description patterns in remote instructions} We derive five recurring language patterns in spatial descriptions (Direct-feature, Relational, Memory, Chained, and Deictic), stemmed from established studies of Language.
\vspace{-1mm}
\item \textbf{Demonstration on collaborative scenarios} We showcase the applications of \sys--Maintenance, Navigation, and Personal assistance, and its potential.
\vspace{-1mm}
\item \textbf{Empirical evaluation} We assess and compare the performance of \sys--cognitive load and usability over an existing remote assistance baseline (voice-only), with implications for integration into existing methods.
\end{itemize}
\vspace{-0.85mm}

By unraveling the ambiguities of verbal interaction in remote assistance and automatically enriching the channel of communication by providing additional visual guidance, \sys advances remote assistance beyond disembodied voice interfaces toward spatially grounded and visually explainable experiences.
\vspace{-1mm}


\section{Related Work}
\subsection{Remote Assistance and Information Sharing}
Remote assistance for procedural tasks has relied on phone or video calls with verbal instructions. While videos provide basic visibility, they lack shared spatial grounding, making it hard to identify precise referents or actions~\cite{druta2021review}. AR and MR research have introduced in-view overlays and annotations to reduce miscommunication~\cite{gurevich2012teleadvisor}. Embedding step-by-step instructions into AR views improved correctness in maintenance tasks~\cite{rebol2021remote, li2025satori}. Surveys further confirm that MR-based remote assistance reduces task errors and increases efficiency~\cite{fidalgo2023survey}. In remote AR guidance, augmenting visual content (e.g., avatar, view-point ray, gestures, pointers) onto the physical realm has shown to facilitate information communication and collaboration~\cite{fink2022re, gronbaek2023partially, zaman2023vicarious, seo2023holomentor, kim2025spaceshare, markov2020effect}. Another set of work shows the use of gestures to enrich data communication--for presentations~\cite{brehmer2025video} as well. Yet, many commercial platforms remain dominated by 2D desktop or mobile interfaces, relying on manual pointers or cursors even with the use of AR, and human-authored annotations for grounding~\cite{druta2021review, wang2025extended, souza2023taxonomy}. 

This motivates \sys to bridge the gap between manual-annotated AR remote assistive technology and conventional speech-only remote guidance. We aim to automatically disambiguate spatial referencing in verbal descriptions using an LLM and a graph-structured reasoning pipeline, enriching a single-channel (verbal) remote guidance to a dual-channel (visual, verbal).

\vspace{-0.85mm}
\subsection{Spatial Referencing in Spoken Language}
\label{subsec:existing_spatial_ref}
Prior works have established that spatial language contains statements that refer to objects with respect to a reference frame~\cite{levelt1993speaking, levelt1982cognitive, shusterman2016frames}, and have shown that people adopt different perspectives and frames of reference--viewer-centered, object-centered, or environment-centered--and that mismatches can cause misunderstandings~\cite{schober1996addressee, brennan1996conceptual, johannsen2013reference}. Taylor and Tversky indicate that spatial descriptions are inherently perspective-dependent~\cite{taylor1996perspective}. In wayfinding scenarios, 
where only the objects near the user are visible, the viewer and the environment-centered (Cardinal directions--North, East, West, South) expressions were used. In contrast, when a scene can be perceived within a single viewpoint, object-centered spatial referencing was more prevalent (e.g., ``Object A to the right of Object B''). Recent interactive and AI systems adopt the object-centered spatial expressions, by pairing language with visual cues to indicate target referents~\cite{kartmann2023interactive, dougan2019learning, chen2022language, howlader2025cora}. Target (the referred object) and Anchor (the figural object used to refer to the Target) attributes (e.g., color, size, shape), relational language, and surrounding visual context further support disambiguation of referents~\cite{dos2015generating, schuz2023rethinking, carbonell2005oral, jadon2024augmented}. These intriguing language patterns of spatial descriptions, and their use cases suggest that explicit spatial anchoring and relational structure play a role in resolving under-specified spoken references.

We motivate our object-centered graph representation from these human spatial description patterns, and use a multimodal LLM to interpret ambiguous target references during remote assistance. As a visual guidance system that situates instructions onto a referred target, we extend a verbally-instructed guidance to visual communication, enabling robust grounding and disambiguation.

\vspace{-0.85mm}
\subsection{Multimodal Cues and Disambiguation}
The seminal work, ``Put-That-There''~\cite{bolt1980put}, established how speech and gesture interrelate each other and treat deictic terms as temporary variables grounded by pointing to spatial targets. Recent systems extend this principle to visualization and XR domains by fusing embodied cues. Han and Issac leverage deictic references (e.g., this, that, here, there) to enrich interaction for visual analytics~\cite{han2024deixis}. GazePointAR employs a gaze and gesture-aware personal assistant to disambiguate under-specified spoken queries in real-time~\cite{lee2024gazepointar}. GesPrompt uses the synchronization between the temporal dimension, speech, and co-speech gestures, to capture richer spatial-temporal intent~\cite{hu2025gesprompt}. Bovo et al. revisits the ``Put-That-There,'' paradigm for XR information placement with scene semantics and head and pointing cues to interpret under-specified commands~\cite{bovo2025revisiting}.

Recent works leverage LLM agents to bind multimodal context (visual, audio, gesture, interaction history) to reason over spatial context~\cite{lee2025walkie, stover2024taggar, zhao2025guided, lee2024gazepointar, han2024deixis, liu2025reality}, and automate grounding. While the fusion of context can mitigate the ambiguity, such techniques assume reliable tracking of gaze, gesture, avatars, or increased sensing computations. Rather than relying on embodied sensing with additional cues, \sys treats speech as the sole instructor channel (lightweight) and resolves ambiguity through a structured referent reasoning pipeline, then generates visual overlays that approximate the disambiguating role of embodied cues.

\vspace{-0.85mm}
\subsection{Intelligent Grounding in XR}
Recent work integrates scene understanding and language models to automate grounding. Guided Reality demonstrates how LLMs and vision models generate visually-enriched task cues embedded into the scene~\cite{zhao2025guided}. Complementary works on dialogue augmentation~\cite{chan2025design} and XR-Objects~\cite{dogan2024augmented}, explore embedding relational semantics from conversation into situated overlays. In parallel, relational scene representations including scene graphs, are increasingly leveraged to capture multi-object relationships and support grounding beyond single entities~\cite{chen2022language, lee2025imaginatear, rosinol20203d, edge2024local, murai2025mast3r, kim2025meta}. These 3D graphs provide hierarchical structure, enabling more transparent relationship predictions than neural representations. ConceptGraphs utilize open-vocabulary detector and scene graphs, enabling systems to query using relational prompts or identify targets~\cite{gu2024conceptgraphs}.

These threads demonstrate the use of multimodal LLM-driven context-aware reasoning. \sys builds on this trajectory, but focuses on the intricacy of verbal descriptions in remote instructions. We ground a speech signal into lightweight, AR anchored visualizations, without requiring additional modalities.

\vspace{-0.85mm}
\subsection{Memory, Recall, and Situatedness}
Visual grounding extends beyond task performance into the domains of long-term memory and recall. Research indicates that pairing speech with visual cues improves information retention and reduces errors. Specifically, Lukianova et al. found that images paired with text in AR instructions, significantly boost recall over text-only conditions. Visual information is processed by the brain more efficiently than linguistic tokens due to their natural semantics over learned symbols~\cite{lukianova2025picture}. Situated visualization studies also show how context-bound representations support recollection and decision-making~\cite{bressa2022data, lee2023design, grubert2016towards}. In AR, spatial markers further support task switching and resumption by visually situating attention and helping users return to a spatial context even after task interruption~\cite{lystbaek2024spatial}. Memory-oriented systems explore how interaction histories can be captured and reused~\cite{shen2024encode, satriadi2023context}. Memoro memorizes prior dialogue to support verbal remembrance nudging~\cite{zulfikar2024memoro}. OmniQuery extends this idea by connecting visual memory with other contextual cues for information search and retrieval~\cite{li2025omniquery}. 

These works suggest that situated persistent memory traces can be used for longer-term usability and spatial reasoning. \sys builds on this by treating every referent as incremental semantic memory. That is, every action and relationship between a user and a referent is stored as a memory. By maintaining an object-centric interaction history within a 3D knowledge graph, \sys can trace prior instructions and context, enabling effective disambiguation of under-specified spatial descriptions.


\vspace{-1mm}
\section{Design of \sys framework}
\vspace{-0.5mm}
\subsection{The Need for Referent Disambiguation in Speech}
\vspace{-0.5mm}
The core challenge in remote assistance is communication clarity. The meaning of an expert's instructions must be quickly understood and acted upon for a collaboration to be successful. However, conventional remote support solutions rely on an asymmetric setup where a local worker streams a visual feed while a remote expert provides speech-only guidance. Since spatial language is inherently ambiguous, this triggers repeated back-and-forth to clarify the intended target and what action should be taken. Recent approaches (e.g., video calls with marking tools or AR tele-assistance) partially address this by enabling experts to add visual annotations, but these cues are typically created manually, adding extra burden.

\sys aims to address this communication bottleneck by designing a disambiguation layer that clarifies the referent behind the remote expert's spoken instruction. We aim to \textit{remove the burden of \textbf{manual visual annotations}} by introducing an automated \textit{\textbf{disambiguation}} pipeline based on speech, and transforming the expert's \textit{\textbf{speech}} into spatially grounded visual indicators.


\subsubsection{Understanding the Language Pattern in Speech Guided Remote Assistance}
\label{subsubsec:formative}
We ground our analysis in established perspectives on spatial reference frames drawn from prior literature: viewer-centered, object-centered, and environment-centered~\cite{levelt1993speaking, levelt1982cognitive, shusterman2016frames, schober1996addressee, brennan1996conceptual, johannsen2013reference, taylor1996perspective}. 
To understand how these perspectives manifest in remote assistance, we examine the recurring linguistic strategies and communication patterns that arise when an expert guides a worker through verbal instructions. We conduct a preliminary formative study of remote verbal instructions with a shared screen view to mimic a remote assistance setting. The insights derived from the study will inform the design of \sys, a framework that disambiguates the spoken spatial references of target objects in remote instructions and produces visual+speech-supported guidance. 

The core aim of the formative study was to analyze the emerging spatial description patterns during remote instructions as the first step, and to base this insight on the design of \sys, and evaluate the effectiveness of our combined approach with AR. 

\subsubsection{Study Setup and Procedure}
We recruited 9 participants (academic researchers, engineers, and students; 8 male, 1 female; aged 27-34; P1-9), paired as instructor (giving directions) and follower (executing them). P1 volunteered to be the designated follower across all sessions (P2-9 being instructors). A 15-minute session with 30 instructions was conducted remotely via Zoom~\cite{zoom} with 2D screen sharing as the shared workspace. The follower was instructed to follow commands literally (dull following), without interpretation, and respond only minimally to the instructor for confirmation. This was to observe and classify spoken patterns of each single turn (a single-trip: query-and-response) conversation, as a multi-turn conversation is a composite that involves more than one single-turn conversations with prior conversation context. The first author (of this research) observed silently, recording transcripts and notable referring expressions. After each session, participants were interviewed about their strategies and difficulties based on the author's session notes. The notes were categorized into high-level themes (\cref{subsubsec:formative_findings}). 

The tasks were designed as an instruction-following activity on a shared desktop view. The 2D screen of the follower was shared with the instructor, representing a remote assistance scenario. For each trial, the instructor was privately informed of the randomly chosen target, ranging from an empty folder to an existing file/icon on the Desktop. The instructor, then, guided the follower to select the item by moving their mouse cursor. To avoid inevitable ambiguity from the non-distinctive feature of a target, the follower was asked, before the study, to create a set of empty folders on their desktop, each named with a unique single letter in the English alphabet. This setup ensured that the instructor always had the freedom to choose any referencing method to refer to a target, without indirectly converging on using a specific referencing method. The first 15 tasks were performed (1) without access to the annotation tool, enforcing speech-only communication, and the other 15 were performed (2) with annotation enabled to examine any shift in participants' strategies upon access to the annotation tool.

\subsubsection{Findings and Implications}
\label{subsubsec:formative_findings}
The study revealed four outstanding patterns of spatial reference, which we thematically coded as Direct Feature, Relational, Memory, and Chained references, following similar groupings of spatial linguistic expressions: Using figural/landmark objects to describe a scene~\cite{taylor1992descriptions, taylor1996perspective, miller1976language}, describing target features~\cite{miller1976language}, using relational/relative descriptions~\cite{garnham1989unified, levinson1996frames}, referring back to previously interacted targets~\cite{Clark1991, levelt1982cognitive, taylor1996perspective}, and deictic references~\cite{levelt1993speaking}. We also observe the use of deictic referencing (e.g., ``that'', ``it'') when the drawing/annotation tool was enabled. We report the total occurrence of each pattern across all tasks, not only the initial descriptions (e.g., ``click the pdf file''), but including the recovery attempts (e.g., ``No, one to the right of it''). One task may involve one or more patterns. The patterns are based on our observation notes (N=187).\\
\vspace{-1.5mm}

\begin{wrapfigure}{r}{0.16\textwidth} 
    \vspace{-10pt}
    \includegraphics[width=0.16\textwidth]{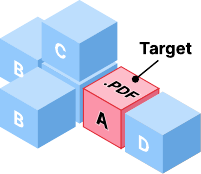}
    \vspace{-15pt}
\end{wrapfigure}
\para{Direct Feature}
The instructor attempted to describe the target directly ($\textit{57.6\%}$) through its intrinsic attributes and features in their initial trials. This includes references to distinct attributes such as the color (``\textit{the \uline{\textbf{red}} file}''), type (``\textit{the \uline{\textbf{PDF}} file}''), or labels (``\textit{the file named \textbf{\uline{A}}}''). Direct Features were effective when the distinguishing attribute was conspicuous. However, when multiple items shared similar features (e.g., similar-colored folders), instructors added additional descriptions, or shifted to other referencing techniques, suggesting that feature-based references provide an important baseline, yet, are fragile for target reference on its own.

\begin{wrapfigure}{r}{0.16\textwidth} 
    \vspace{-10pt}
    \includegraphics[width=0.16\textwidth]{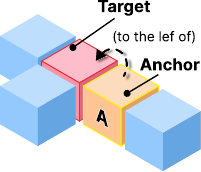}
    \vspace{-15pt}
\end{wrapfigure}
\para{Relational}
When Direct Feature referencing was insufficient, the participants used this type ($\textit{31.2\%}$). It uses an anchor referent/object that is more conspicuous and acts as a landmark to the target. Phrases such as ``\textit{the one \textbf{\textit{\uline{to the left of the}}}} \uline{\textit{yellow file}}.'' Relational referencing allowed instructors to disambiguate targets even when features overlapped, but it required the follower to correctly identify the anchor object first. When the follower still failed to identify the anchor, the instructor resorted back to guiding the follower's mouse cursor as the anchor point.
For example, ``\textit{move a little more to the left}'' or ``\textit{to your right}'' (micro-guidance). The prevalence of this pattern indicates the importance of modeling spatial relations.

\begin{wrapfigure}{r}{0.16\textwidth} 
    \vspace{-10pt}
    \includegraphics[width=0.16\textwidth]{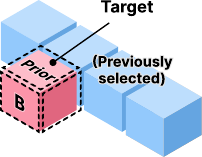}
    \vspace{-15pt}
\end{wrapfigure}

\para{Memory}
Participants used prior knowledge to indicate an anchor referent (a distinct referent to indicate the target) ($\textit{11.2\%}$). This pattern refers back to the objects that had been mentioned or manipulated earlier in the session, the short-term shared knowledge between the instructor and the follower. An example phrase could be: ``\textit{the file we \textbf{\textit{\uline{previously}}} selected}''. This strategy highlights how the anchor point in referring a target object is not limited to an object's distinct attributes, but can also be based on the shared experience the two collaborators engaged in. This shared experience then becomes a resource for grounding cues as well. However, one of the participants indicated that the Memory pattern caused them additional cognitive overhead, as they tried to recall their interaction history. This referencing motivates our system design to retain interaction history as part of the grounding step.
\begin{wrapfigure}{r}{0.16\textwidth} 
    \vspace{-10pt}
    \includegraphics[width=0.16\textwidth]{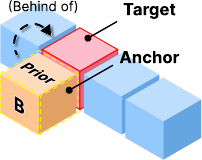}
    \vspace{-15pt}
\end{wrapfigure}

\para{Chained}
This is a composite pattern. In a single utterance, it layers Direct Feature, Relational, and/or Memory-based cues. For instance: ``\textit{the folder \uline{\textit{\textbf{behind}}} the one we \uline{\textit{selected earlier}}}'' (Relational, Memory). These references emerged when the description complexity of the anchor or target increased, or when initial guidance attempts failed. While effective, an elongated chain of instructions was reported to frustrate both the follower and the instructor, often leading them to merely follow mouse cursor-based micro-guidance. The Chained referencing pattern illustrates the need for a compositional representation such as a graph that can integrate cues beyond spatial relations between referents. Its statistics are broken down into its individual patterns.

\begin{wrapfigure}{r}{0.16\textwidth} 
    \vspace{-10pt}
    \includegraphics[width=0.16\textwidth]{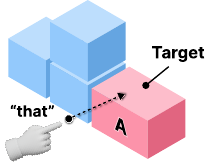}
    \vspace{-15pt}
\end{wrapfigure}
\para{Deictic}
This pattern was the most prominently used with the visual annotation tool. For example, ``\textit{\textbf{\uline{that}} one}'', ``\textit{\textbf{\uline{it}}}'' (while visually marking the target). However, this pattern is dependent on additional cue such as visual indication or explicit gesture pointing~\cite{bolt1980put, lee2024gazepointar, hu2025gesprompt}. As we disambiguate solely on speech cue, we consider this pattern as out of our scope, and discuss in our future work discussion (\cref{sec:takeaways_and_discussion}).

\subsection{Design Rationale}
Our formative study validated that the linguistic spatial patterns established in (\cref{subsec:existing_spatial_ref}) also emerge in remote verbal guidance scenarios. These descriptions relied on (1) visual features or attributes of objects, (2) relations among entities, and (3) shared interaction context. We use a graph representation that retains these patterns and supports structured interpretation and reasoning over referred targets in spoken instructions. Leveraging this representation, \textit{we disambiguate the referred target in speech instructions} of remote assistance and project AR guidance onto the resolved referent for \textit{explicit visual guidance}. We detail the design considerations for our framework in the following:
\vspace{2mm}

\noindent
\dc{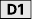} \para{Spatial representation for referent resolution}
\vspace{-1mm}
A referent-centered 3D spatial representation that can retain its own attributes (e.g., 6DoF transformation, color, size, shape), and provide a way to connect and traverse other neighboring referents. The structure must be able to represent the three spatial expressions (Direct, Relational, and Chained).\\

\vspace{-2.5mm}
\noindent
\dc{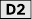} \para{Retain prior interaction context} 
\vspace{-1mm}
The spatial representation not only retains what attributes an entity/referent has, but also the temporal grounding (interaction history--Memory). The system must be able to retrieve the prior activities applied to each referent in a form of an episodic memory.\\

\vspace{-2.5mm}
\noindent
\dc{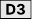} \para{End-to-end generation of visual indicator}
\vspace{-1mm}
To address the adoption barrier of manual annotation, the system must convert referential language into a concrete spatial output automatically. This requires an end-to-end pipeline that maps utterances to candidate referents and emits an explicit visual indication, reducing reliance on repeated verbal clarification.\\

\vspace{-2.5mm}
\noindent
\dc{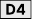} \para{Lightweight instrumentation}
\vspace{-1mm}
To remain practical across diverse AR deployment settings, the referent disambiguation from a verbal instruction, must not require any additional embodied cues (e.g., gaze, gesture) or hardware-specific dependencies, while capturing comprehensive context about the user (e.g., space, time, activity, intent, referent)~\cite{kim2025explainable, jang2005unified, lee2019remote}.\\

In designing \sys, we treat speech as an expressive specification of referential intent, and convert it into automatic anchored visual guidance to make spatial grounding attainable in lightweight settings without additional cues such as gaze or gestures. We aim to disambiguate a target from verbal expressions, into visually grounded guidance using graph-based reasoning with an LLM and AR, enriching remote communication beyond manual visual annotations with verbal instructions.

\begin{figure*}[!t]
    \centering
    \includegraphics[width=\linewidth]{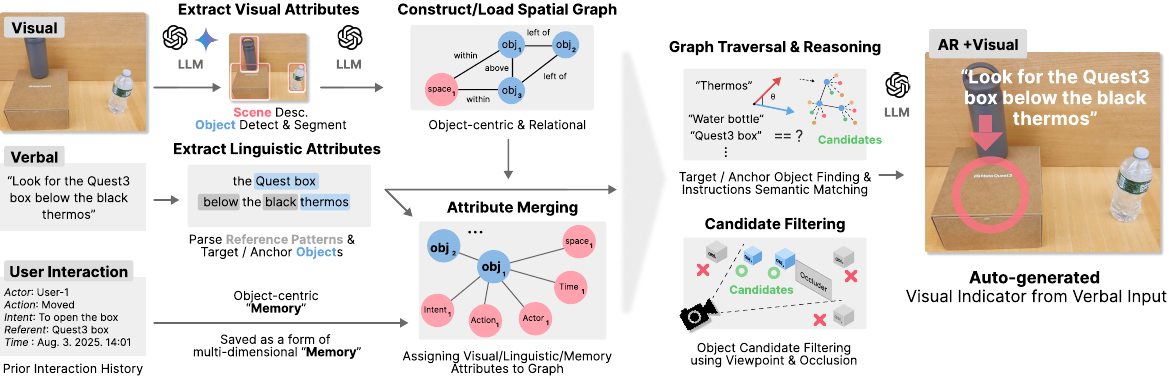}
    \vspace{-7.25mm}
    \caption{
    End-to-end pipeline of \sys: From speech with visual inputs and prior memories (if present), \sys extracts the visible, linguistic attributes, associates them with a relational graph representation, and generates an AR visual indicator.
    }
    \vspace{-1.25mm}
    \label{fig:pipeline}
\vspace{-4mm}
\end{figure*}

\subsection{System Implementation}
\para{System Overview}
\sys follows an end-to-end workflow: (1) spoken instructions are transcribed and interpreted; (2) the target, anchor (if applicable), and the descriptions of each object are parsed; (3) an object-centric graph is constructed from the visible objects and referents (\dc{figures/DC1.pdf}); (4) attributes per node are assigned from the visible features; (5) the referred object is assigned with its requested interaction history (\dc{figures/DC2.pdf}); and (6) an augmented visual overlay is situated onto the physical environment (\dc{figures/DC3.pdf}), without necessitating other cues beyond speech (\dc{figures/DC4.pdf}).

At the core of \sys lies the object-centric graph. It allows every visible physical object to retain multi-dimensional attributes--space, time, action, intent, and actor--capturing the history of interactions, while also encoding its own feature descriptors (e.g., color, shape), and spatial relations between neighboring objects. This persistent object-based representation supports the integration of the four referencing patterns and facilitates the disambiguation of object references during remote verbal communication. The pipeline is illustrated in \cref{fig:pipeline}. Please refer to our Supplementary Material for more implementation details.

\para{Implementation Environment}
We offload the computation overhead from the client's device (via a client-server architecture). On the client's device, Unity AR Foundation was used, and we use a custom Python server for the reasoning backend. Unity handles the visual capture, voice recording, 6DoF pose extraction of objects, and AR anchoring of contents. Objects are represented internally as JSON nodes before being incorporated into the object-centric graph on the server. Speech is transcribed using Whisper~\cite{whisperUnity}, and utterance parsing, reasoning, and resolution are handled by \gpt, followed by the text-embedding-3-small model, which generates the embeddings for object attributes. \gemini was used to localize, segment, and classify objects, and perform visual analysis of the scene. Embeddings of object attributes are cached and reused for optimality.
\vspace{-1mm}

\subsubsection{Remote Instruction Parsing and Attribute Extraction}
\vspace{-0.5mm}
When \sys receives a spoken instruction, it parses the utterance into a pre-defined structure with an LLM. The structure encodes the linguistic dimensions: $\{target object\}$, one or more $\{anchor object\}$, their respective object $\{class/label\}$, $\{description/features\}$, a $\{relational phrase\}$, and any $\{action\}$, $\{intent\}$, or $\{temporal\}$ cues. As shown in \cref{fig:attribute_extraction}, ``\textit{The black thermos above the Quest 3 box, next to the Poland spring water we talked about a minute ago}'' yields the target node (``\textit{thermos}''), the anchor node (``\textit{box}''), associated with the attributes of each node, and their Direct Feature, Relational, and Memory-based relationship. Generic nouns such as ``\textit{thing}'', or ``\textit{it}'' are not treated as a target label or description, but rather as the question of the referent. 

\vspace{-1mm}
\subsubsection{Object-centric Relational Graph Construction}
\begin{figure}[!t]
    \centering
    \includegraphics[width=\linewidth]{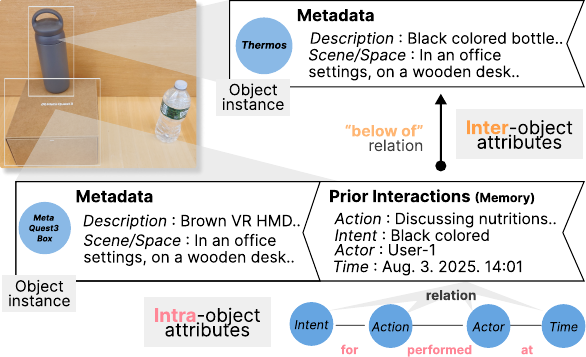}
    \vspace{-6mm}
    \caption{Object-centric relational graph: Each object maintains a graph representation that holds \textcolor{object_red}{\textbf{Intra-object attributes}}--Space, Action, Intent, Actor, Time, and Metadata--and is connected with other objects nodes--\textcolor{object_orange}{\textbf{Inter-object attributes}}.}
    \vspace{-2mm}
    \label{fig:object_centric_graph}
\vspace{-3.5mm}
\end{figure}

\vspace{-0.1mm}
Upon the first encounter with a scene and the localization of an object, we build a relational graph, mapping the logical node-object to the physical object (\cref{fig:object_centric_graph}). The graph is retained across AR sessions enabling users to permanently refer back to their prior interactions with the objects. \sys bypasses the graph construction step once it already holds a representation of the scene.

\para{Object Detection and Segmentation}
We localize the visible objects in the captured snapshot via a vision-capable LLM. It is prompted to localize all visible objects, as well as to extract their descriptive features (e.g., color, class-label, shape). Specifically, we guide the LLM to localize the parsed target and anchor objects. 

\para{Graph Construction}
Once objects in a scene are localized, and their features are extracted, a relational graph is constructed. Every localized physical object is registered as a logical node in the graph, and the nodes are connected by spatial relationships between their neighboring nodes. The spatial graph is constructed using the six spatial relational properties: ``\textit{left}'', ``\textit{right}'', ``\textit{above}'', ``\textit{below}'', ``\textit{in-front-of}'', and ``\textit{behind-of}''. We define an object to be ``in relation'' to another only when it is within a half-meter radius (r=50cm). The 3D center point of a localized node, derived from the axis-aligned 3D bounding box of a target object, is used to determine the spatial relation between objects, for simplicity. 

\para{Assigning Attributes to Graph Nodes}
We assign identified features (e.g., color, class-label) of each object extracted from earlier steps, to logical node attributes and its spatial context (the description of the overall scene the objects are in; e.g., ``\textit{Desk with laptop and coffee on the side.}''). Then, we initialize each graph node with an empty memory field (\cref{fig:attribute_extraction}), which binds the temporal footprint and action/intent history of user(s) when an interaction occurs.

The ontology-driven structures~\cite{kim2017ontology, murai2025mast3r, edge2024local, gu2024conceptgraphs} have shown that graphs can preserve explicit relations and histories, ensuring that references are resolved through paths that remain interpretable and explainable, unlike flat vector-based representations. By combining the concept of a scene graph with multi-context--space, time, actor, action, intent~\cite{kim2025explainable, jang2005unified, shakeri2023user}--we represent spatial relationships across objects as well as their interaction histories and context.

\vspace{-1mm}
\subsubsection{Referent Inference and Reasoning}
\vspace{-1.3mm}

\para{Compositional and Chained Reasoning}
\sys resolves utterances that involve multiple dimensions of reference. Each node in the object-centric graph consists of not only what the object is, but also where it is, what has been done to it, and when; on top of the spatial relationship between its neighboring objects. With the integration of relational, spatial, temporal, action, and intent-based dimensions, the graph supports compositional reasoning across multiple anchors and chained references. For instance, ``\textit{the cube behind the sphere and in front of the machine}'' can be resolved by intersecting relational paths from the two anchor nodes, while ``\textit{the bolt next to the panel we fixed earlier}'' requires traversing both spatial and temporal-action histories. 

\begin{figure}[!t]
    \centering
    \includegraphics[width=\linewidth]{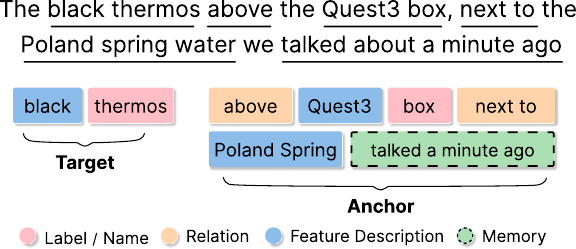}
    \vspace{-6mm}
    \caption{Attribute parsing: Transcribed text of verbal instructions is extracted into a structure via LLM.}
    \vspace{-2mm}
    \label{fig:attribute_extraction}
\vspace{-3mm}
\end{figure}

\para{Semantic Embeddings and Attribute Matching}
Relational properties are fixed to six terms (e.g., two of which are ``\textit{left}'' and  ``\textit{right}''). However, Direct Feature (`\textit{label}', `\textit{description}' in \cref{fig:attribute_extraction}) or Memory-based referencing rely on descriptive language, not constrained to a set of pre-defined vocabularies. To handle this variability, \sys computes semantic similarity between each graph node and the parsed attributes of the target/anchors (cosine similarity between vectors). We select the top five candidate nodes ($k=5$) with highest attribute description similarity, while satisfying relational and memory attribute alignment. Then, we perform an LLM-based reasoning to pinpoint the referred node. The first-pass semantic similarity node filtering, not only reduces the input context counts passed to the LLM in the final step, but also provides robustness to linguistic variation in referents beyond Bag-of-Words or naive keyword matching.

\para{Viewpoint-aware Candidate Filtering}
\sys employs object-level frustum and occlusion-culling of object nodes, to minimize the similarity computation checks of candidate nodes in the graph. Similar to the culling techniques in graphics engines, only the nodes that are in view are retained, while those outside are discarded. Then, naive occlusion culling is performed by casting an AR ray (physical surface depth checks) to the known object-node position, for depth-testing following the pseudo-code:
\vspace{-5mm}
\begin{center}
\[
    \text{if } (D_A > D_B + \Delta_{scale}) \;\; \text{then: ${occluded}$}
    \vspace{-1.25mm}
\]
{
\vspace{-1mm}
\small
(where $A = {{cam\_to\_obj\_dist}}$, \;\; $B = {{ray\_hit\_dist}}$,
\par $\Delta_{scale} = {target\_obj\_scale}$)
}
\end{center}
\vspace{-0.5mm}

\noindent
At the end of the reasoning pipeline, we employ a fallback mechanism in the case of conflicting target referencing or reasoning failures. We leverage an evaluation LLM agent to verify whether the final candidate referent satisfies all the conditions without ambiguities. Upon conflict, it falls back to the raw transcription anchoring, instead of a visual pointer to avoid any uninformed guidance.

\subsubsection{Capturing Interaction History}
\vspace{-1.75mm}
\para{Retaining Action Footprints}
A central principle of \sys is that memory is not an auxiliary log but an integral part of the graph itself. Each node accumulates a persistent history of interactions: the actions taken (e.g., ``\textit{moved}'', ``\textit{rotated}''), the actors involved, and the temporal footprint of those actions. By storing these directly onto object-nodes, \sys ensures that references such as ``the panel we looked at yesterday'' or ``the folder we opened earlier'' can be naturally resolved by traversing temporal-action attributes. This approach turns every object into a site of accumulated memory, anchoring its evolving state across time.

\para{Revisability of Object-nodes}
The object nodes of the graph are preserved across sessions. However, the attribute of an object can be updated through a user interaction. When an object is transformed, moved, or altered, \sys updates its attributes in place. This interaction `Action' is recorded in the interaction history of the object-node, and is appended to its existing interaction data. This supports the Memory-based referencing pattern, enabling the recall of a referent based on prior interactions.

\vspace{-1mm}
\subsubsection{Anchoring and Visualizing Indicator}
\vspace{-1.5mm}
\para{AR Visual Indicator Anchoring}
Once a referent is identified through graph traversal reasoning, \sys generates an AR visual indicator to situate the instruction in the physical environment. We anchor a directional arrow pointer directly above the identified object-node, ensuring that the referent is immediately visible to the remote user. This anchoring is persistent until an action (e.g., ``\textit{moving}'') has been performed to the referent.

\para{Instruction Summarization and Step Ordering}
On top of the visual indicator, \sys overlays an AR instruction panel generated by an LLM. The utterance is summarized into a concise action description, such as``\textit{tighten the bolt}'' or ``\textit{open the left panel}''. When multiple steps are parsed from the instruction sequence, the system presents them in an alphabetically ordered list (e.g., A,B,C), guiding the user through the required operations in order. This summarization clarifies the required action with intent, reducing the cognitive burden of parsing long-utterance tasks.

\para{Misguidance Avoidance}
In cases where \sys fails to interpret the instructions--involving conflicting conditions or loss of nuances in summarization, \sys alternatively displays the raw transcription of the spoken instruction without the auto-generated visual pointer, as a safe fallback mechanism. This enables \sys to avoid presenting any misleading visual pointers, passing control over to the users for their interpretation (maintaining user-agency). This approach preserves fidelity to the original input while ensuring that users retain control when automation falls short. This design maintains the system fail-safe.
\vspace{1.25mm}

Anchoring a visual indicator completes our ``Speech-to-Graph-to-Overlay'' pipeline. The arrow indicator provides spatial grounding, while the textual overlay delivers actionable guidance. Together, they transform remote verbal instructions into a spatial indicator that disambiguates the object-of-interest and clarifies actions.

\begin{figure*}[!t]
    \centering
    \includegraphics[width=\linewidth]{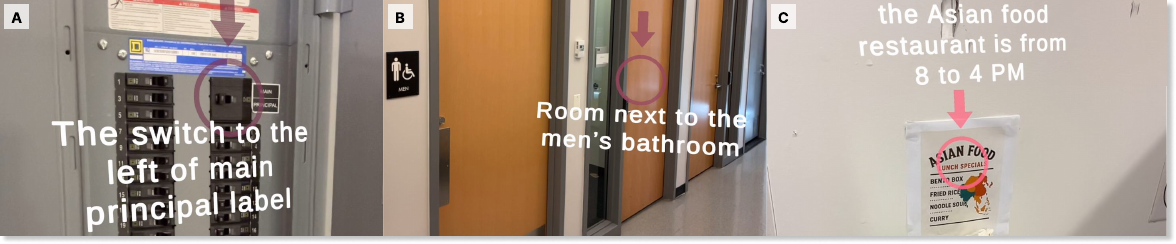}
    \vspace{-7mm}
    \caption{
    Three use case scenarios of \sys. \textbf{(A)} Remote Maintenance; \textbf{(B)} Indoor Navigation; and \textbf{(C)} Personal AI assistant.
    }
    \vspace{-2mm}
    \label{fig:usecase_application}
\vspace{-4mm}
\end{figure*}

\section{Use Cases and Applications}
\vspace{-0.25mm}
We illustrate how \sys operates across the applications of Remote maintenance, Navigation, and Personal assistance, emphasizing its ability to provide clarity in conversations, preserve context, and reduce ambiguity in the topic of discussion. Each case demonstrates how referent disambiguation can transform under-specified speech queries into spatially guided instructions (\cref{fig:usecase_application}).

\subsection{Case Study 1: Speech-based Visual Annotation}
In modern remote collaborative maintenance platforms (e.g., TeamViewer Assist AR~\cite{teamviewerassist}, or Microsoft Dynamics 365 Remote Assist/Teams~\cite{msremoteassist}), a remote expert supports the technician in the field primarily through spoken instructions while observing a shared video feed. Other platforms provide annotation tools that allow experts to label regions in the view. Yet, verbal description is the primary source of assistance, and visual annotation requires a human expert to manually indicate the target as well. With \sys, an utterance (e.g., ``\textit{locate the second fuse from the left, just below the green wire}'') is parsed into an object-centric graph, and the remote instructor's referent is automatically disambiguated and indicated with a visual overlay. This transformation provides the technician with unambiguous, visual directions, without requiring manual effort to visually indicate the referred target.

\subsection{Case Study 2: Mapping the Speech to Visual Map}
When describing the route to a destination, directions are conveyed step by step, combining landmarks and relational anchors (``\textit{walk straight until you see the entrance, and turn left}''). Such instructions require the listener to maintain a mental drawing of the route, assembling each fragment into an imagined plan of the environment. While natural in daily conversations, this approach can be difficult to follow in unfamiliar or complex spaces, where the accuracy of remembering each segment is critical. \sys can convert the verbal directions into a situated visual guidance using AR. Instead of relying on the user's ``mental map,'' from the verbal explanation, users can visualize the directions. The wayfinding experience becomes easier to follow and transparent. The overlay makes clear what the speaker meant by ``\textit{entrance}'' or ``\textit{left},'' reducing the likelihood of misunderstanding, in spoken instructions.

\subsection{Case Study 3: Disambiguating the Query-of-interest}
\sys is useful in verbally conversing with a personal AI assistant (e.g., Gemini) as well. Here, the spoken communication can introduce ambiguity in identifying the intended referent. The user may refer to Referent A, but the AI may misinterpret it as Referent B, providing an incorrect response to the user's query. \sys can mitigate this by providing a visual indicator (e.g., a dot indicating the object-of-interest ~\cite{lee2024gazepointar}) after receiving the speech input. Furthermore, this capability suggests a potential pathway towards the transformation of a customer-service AI chat-bot into a visually grounded agent that can visually guide (on top of speech/textual) even with a single shared snapshot capture.


\vspace{-0.5mm}
\section{Evaluation}
We evaluate \sys as a \textit{referent disambiguation mechanism} that converts verbal remote instructions into spatially grounded AR guidance, by resolving the intended referent and providing a visual indicator. Our evaluation consists of two parts. (1) End-to-end \textbf{impact}: we first test \textit{whether grounding speech-only instructions into AR guidance with \sys, \uline{improves user performance} compared to a conventional voice-only baseline} (\cref{subsec:user_eval}) and (2) the \textbf{feasibility} of our mechanism: we analyze the spoken instructions and interaction traces logged during our open-ended study to quantify \textit{\uline{how reliable our mechanism resolves the intended referent}} (\cref{subsubsec:open_ended}).

\subsection{User Evaluation: Quantifying the Impact}
\label{subsec:user_eval}
\subsubsection{Study Setup and Procedure}
\para{Participants}
We recruited 18 participants (12 male, 6 female, aged 22-34). All participants were either Full Professional Proficiency (N=4) or Native (N=14) in English. The 7-point Likert scale indicating prior familiarity with XR systems (1:None; 7:Experienced) varied widely ($\mu$=2.7, $\sigma$=1.2), but the participants were not exposed to the task or conditions prior to the experiment. The participants (P1-P18) were compensated with a \$15 gift card.

\para{Apparatus}
Participants were given a Meta Quest3 MR HMD (passthrough mode) with its controller, to participate in the study. The HMD was to provide a controlled testbed across participants that ruled out any arm fatigue concerns (e.g., Gorilla Arm Effect) that may arise from a hand-held AR device, and allow users to perform tasks with hands (hands-free), under a practical remote assistance workflow (``receive-instructions-and-execute''). The spatial layout of the physical scene and the eight virtual cubes (mock physical referents) were registered and anchored prior to the study. To maintain uniform study conditions, the coordinate system was synchronized across sessions, and the study was conducted on the same desk settings in a lab (refer to \suppl). 

\para{Procedure and Measures}
Participants were given a cube interaction task under three conditions with different modalities (each modality is labeled, a `block'): (1) {\textit{Audio}} (spoken instructions only; denoted ``\textbf{\audioc}''); (2) {\textit{Audio+Visual Indicator+Full transcription}} (verbatim instruction displayed; denoted ``\textbf{\fullc}''); (3) {\textit{Audio+Visual Indicator+Summarized transcription}} (condensed directive above a target referent; denoted ``\textbf{\summaryc}''). Note that the speech-only condition (1--\audioc) is our \textbf{\uline{baseline}}, and 2--\fullc and 3--\summaryc, use \uline{\sys} to disambiguate instructions and provide spatial guidance (generated before the study). Within each block, participants complete four sub-blocks consisting of three trials, each covering one of four spatial referencing patterns (\cref{subsubsec:formative}).

Every user action was logged to derive task completion time, accuracy, and interaction traces. After each sub-block, participants filled out a 0-100 scaled RTLX (Raw TLX; Unweighted NASA-TLX) and a Single Ease Question (SEQ). After each modality block, they completed another RTLX for overall experience assessment. A post-study questionnaire was given to collect overall user experience, followed by a semi-structured interview. The study conditions and tasks were presented in a counterbalanced order to mitigate sequence effects. Also, to maintain consistent motor factors (pinch, hand gestures), participants were instructed to start the task at the same designated location, and the target cubes were repositioned to maintain an equal distance of 35cm ($Distance_{hand\_to\_target}$). All instructions were delivered as AI-synthesized speech to avoid between-speaker prosody effects, and the audio was played at the beginning of every trial. When the sub-block referencing pattern type is ``Memory-based,'' users were shown to remember that this is ``\textit{the cube referred by the memory}''. Each of the eight cubes was textured uniquely. We collected a total of 1,296 trials across all participants (3 blocks × 3 trials × 4 sub-blocks × 2 task types × 18 participants), excluding a 20-minute functionality familiarization phase (tutorial).

\para{Tasks}
The tasks simulated remote guidance scenarios, where participants were verbally guided to select a target cube among distractors, with instructions balanced across reference styles (each trial is given a single task). Each participant was given equal aggregated counts of trials and tasks:

\vspace{-3mm}
\begin{itemize}
    \item \parac{Locate} 
    Identify and select a target cube among distractors by hitting it with a virtual hammer mapped to the Quest controller. The task hypothesizes a ``find'' instruction, given a description (e.g., ``\textit{Locate the purple striped cube}'').
    \vspace{-2.5mm}
    \item \parac{Move} 
    Move a specified cube from its current location to a designated target position on the desk. The action is performed using a pinching gesture (bare-handed). The scenario hypothesizes a remote-assisted task that involves transformation of an on-site object. The task is considered complete once a cube was selected and moved to any position (e.g., ``\textit{Move the red cube to the left of the blue dotted cube}'').
\end{itemize}
\vspace{-2mm}

\noindent
In \audioc, instructions were delivered verbally with no additional cues mimicking traditional speech-only remote guidance. In \fullc, the spoken instruction was displayed verbatim above the target with an arrow. For \summaryc, the instruction was condensed into a concise directive and displayed alongside the arrow. We conduct a within-subject study with these hypotheses: \textbf{(H1)} \sys will disambiguate instructions and improve clarity than the baseline; \textbf{(H2)} Summarized instructions will be the top choice; \textbf{(H3)} The advanced task (Move) will have higher demand for disambiguation.

\vspace{-0.75mm}
\subsubsection{Results}
\label{subsec:result}
\vspace{-0.75mm}
We report our findings at three levels computed for each Task: Overall effects across all trials, Modality/block-level comparisons, and Effects on the performance for each referencing pattern. For each block, we perform a repeated measures ANOVA test (when normality held), or Friedman followed by Bonferroni-corrected pairwise tests (or Wilcoxon signed-rank) with effect sizes. We use Shapiro-Wilk for normality checks. For each section and task, we summarize the insights, and provide statistical grounds behind them (as the tasks involve different levels of complexity and motor).

\begin{figure}[!t]
    \centering
    \includegraphics[width=\linewidth]{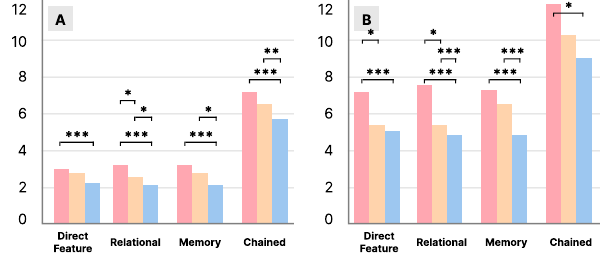}
    \vspace{-6mm}
    \caption{Comparison of median task completion time per referencing pattern : \textbf{(A)} Locate Task, \textbf{(B)} Move Task result; *** $p<.001$, ** $p<.01$, * $p<.05$; \textcolor{plot_red}{\textbf{\audioc}},
    \textcolor{plot_orange}{\textbf{\fullc}},
    \textcolor{plot_blue}{\textbf{\summaryc}}.}
    \vspace{-1.5mm}
    \label{fig:bar_plot}
\vspace{-4mm}
\end{figure}

\para{Task Completion Time and Accuracy}
In both tasks, the disambiguated-instruction conditions (\fullc, \summaryc) reduced completion time without harming accuracy compared to \audioc. The accuracy of Locate was near maximum under all conditions with only the visual guidance time varying. For Move, they significantly improved both speed and accuracy, with \summaryc providing the strongest benefits, and \fullc not showing a strong trend with \audioc.

\textit{Locate:}
the median completion time (ascending sorted) for the conditions was: 
\summaryc $3.25(\pm0.53)s$, 
\fullc $4.08(\pm0.40)s$, and
\audioc $4.33(\pm0.41)s$. Friedman test showed significance across conditions on Time (${\chi^2=21.00}$, ${p<.001}$), and pairwise comparison indicated: $\summaryc{<}\audioc$ (${p<.001}$), $\summaryc{<}\fullc$ (${p<.001}$). $\fullc_{vs}\audioc$ did not show any significance. Accuracy was:
\summaryc $0.998(\pm0.001)$, 
\fullc $0.991(\pm0.027)$, and
\audioc $0.986(\pm0.032)$.
No significance was found among conditions. 

\textit{Move:}
this task requires more complex reasoning (identify and move). The mean completion time were: 
\summaryc $6.33 (\pm1.07)s$,
\fullc $7.94(\pm1.22)s$, and
\audioc $9.31(\pm2.28)s$. Conditions significantly affected time (${\chi^2=20.33}$, ${p<.001}$), with $\summaryc{<}\fullc{<}\audioc$ (${p<.05}$). Accuracy of \fullc and \summaryc also improved over \audioc:
\summaryc $0.731(\pm0.061)$, 
\fullc $0.676(\pm0.137)$, and
\audioc $0.644(\pm0.130)$, (${\chi^2=8.04}$, ${p<.018}$). A Wilcoxon signed-rank test showed: $\summaryc{>}\audioc$ (${p<.020}$), while no other conditions were found to be significant. This indicates that \summaryc improves both speed and accuracy.

\para{Perceived Difficulty}
Disambiguated guidance (ours) reduced perceived difficulty (higher: more difficult) robustly for the Move task, while the Locate task showed non-significance. Difficulty was measured with a 0-to-100 scale (0:Low; 100:High).

\textit{Locate:}
the difficulty of each condition in ascending order:
\summaryc $26.39(\pm23.01)$, 
\fullc $28.19(\pm24.04)$, and
\audioc $30.97(\pm19.61)$, did not reap any significance.

\textit{Move:}
showed significance (RM-ANOVA ${F(2,34)=7.85}$, ${p<.05}$) with $\fullc_{vs}\audioc$ (post-hoc ${p=.020}$), and $\summaryc_{vs}\audioc$ (post-hoc ${p=.016}$). The difficulty of each condition was:
\summaryc $33.47(\pm24.54)$, 
\fullc $34.72(\pm23.13)$, and
\audioc $45.14(\pm17.60)$.

\para{Perceived Confidence}
The perceived confidence in answers increased with \summaryc significantly higher than \audioc, in Move. Locate did not exhibit any significant trend. Confidence was measured with a 0-to-100 scale (0:Low; 100:High).

\textit{Locate:}
No significance among conditions was found. Each condition scored:
\fullc $89.17(\pm13.09)$
\summaryc $88.89(\pm11.48)$, and
\audioc $81.94(\pm18.70)$.

\textit{Move:}
Confidence in the answer of each condition was:
\summaryc $88.89(\pm11.12)$
\fullc $85.56(\pm15.30)$, and
\audioc $77.64(\pm20.89)$. Showing significance (RM-ANOVA ${F_{(2,34)}=5.61}$, ${p<.01}$) for $\summaryc_{vs}\audioc$ (post-hoc ${p=.036}$).

\para{Cognitive Workload}
Interestingly, \fullc had lower average load than \summaryc in Mental Load and Effort, across both tasks. While it did not reap any significance between \summaryc and \fullc, \fullc showed lower mean than even \summaryc. Only for the Move task did both the spatial guidance (\fullc, \summaryc) reduced mental demand as well as effort, relative to \audioc. The Locate task did not show any statistically meaningful pattern. Each load was measured with a 0-to-100 scale (0:Low; 100:High). 

\textit{Locate:}
Mental demand and Effort were:
\fullc $23.19(\pm23.62)$, \summaryc $27.36(\pm25.60)$ , \audioc $30.69(\pm20.25)$, and 
\fullc $22.08(\pm19.50)$, \summaryc $25.97(\pm22.48)$, \audioc $29.31(\pm18.49)$.

\textit{Move:}
Mental demand and Effort were:
\fullc $32.22(\pm24.16)$, \summaryc $34.31(\pm27.93)$, \audioc $47.22(\pm20.45)$, and 
\fullc $33.06(\pm22.70)$, \summaryc $34.31(\pm23.79)$, \audioc $46.53(\pm17.62)$, respectively.
It shows significance (RM-ANOVA ${F_{(2,34)}=7.80}$, ${p<.01}$) on $\fullc_{vs}\audioc$ (post-hoc ${p=.006}$) for Mental demand, and Effort also shows significance (RM-ANOVA ${F_{(2,34)}=9.72}$, ${p<.001}$) in $\fullc_{vs}\audioc$ (post-hoc ${p=.005}$), and $\summaryc_{vs}\audioc$ (post-hoc ${p=.013}$).

\para{Effects Per Referencing Pattern}
We examine how referencing patterns (Direct Feature, Relational, Memory, Chained) shape performance within each task (Locate, Move) and by modality (\audioc, \fullc, \summaryc). \summaryc consistently shortened completion time relative to \audioc across all patterns and tasks, and it also showed better results than \fullc for Relational, Memory, and Chained speech patterns. The effect is largest whenever the utterance requires cross-object reasoning or recall (Memory, Chained), where the concise directive appears to reduce parsing effort while the arrow removes residual spatial ambiguity (\cref{fig:bar_plot}). For Accuracy, Locate does not show a noticeable pattern, while for the Move task, \summaryc yields the most gains for Memory-based references. Please refer to Supplementary Material for more analysis.

\para{Usability and Preference}
Participants indicated the value of disambiguated guidance positively, rating (7-point scale) the use of \summaryc, at $5.37(\pm1.26)$ with 79\% rated higher than or equal to 5, and 53\% rated higher or equal to 6. The visual reliance compared to \audioc was ${6.84}$ vs. ${3.16}$ (ratio). The participants indicated that the visual anchoring reduces memory burden and expedites action ``\textit{Seeing the arrows made it easier to remember the steps to be taken}'' (P3). When asked to rank the preferred mode of assistance, \summaryc was ranked highest (N=11), followed by \fullc(N=6), and \audioc (N=1). ``\textit{\summaryc captures key points without overwhelming detail, and the arrow makes instructions easy to follow}'' (P5). The \audioc, which is the most common way of remote assistance in the field, was viewed as insufficient by the participants. ``\textit{\audioc was less helpful I need to think and remember details}'' (P11); ``\textit{After getting used to the visual one, audio was just distracting.}'' (P8). 

\subsection{System Evaluation: Gauging the Feasibility}
\label{subsubsec:open_ended}
With the same participants post-study, we assess our framework as a whole with the participants' own verbal input. In this session, the participants were not constrained to any specific tasks, but acted as both the remote expert (who provided the verbal instructions) and the local user (who shared the view). An example query can be: ``\textit{Locate the phone to the left of the laptop}'' (while having a phone and a laptop in sight). This session was conducted to gauge the raw technical capabilities of our framework. The participants were instructed to provide any verbal queries and provide open-ended feedback on their user experience of \sys. 

We collected 81 user query instances along with qualitative feedback from the participants. Our analysis indicates that \sys successfully processed $\textit{77.8\%}$ (counts: 63) of total queries, while $\textit{13.6\%}$ (11) triggered our fall-back mechanism (\fullc). The verbal input contained incorrectly parsed noise or filler utterances (e.g., ``um..'') leading to $\textit{8.6\%}$ (7) speech recognition errors. Overall, the system was commented on positively on its ability to disambiguate instructions through \fullc or \summaryc visualizations (P1,P3,P4,P9,P10). However, one participant expressed frustration with the persistent visualization of \fullc transcriptions (``\textit{I don't know if it's the system or me}'') (P1). The fail-safe mechanism where the raw transcription is provided to the user upon conflicts or failures caused confusion between the system's reliability and voice recognition errors. Another participant perceived the same fall-back behavior as beneficial, noting ``\textit{fallback helped, it shouldn't just fail, at least it shows where it possibly went wrong.}'' (P9). \sys was unable to correctly parse queries involving unsupported spatial expressions. For example, multi-step spatial queries (multiple Chaining spatial references): ``\textit{second to the right of}'', ``\textit{in between object A and B}'' (P2), and ``\textit{what's next to the one on the right}'' (P7). Also, a user view-oriented explanation (``\textit{one to the right of what I'm viewing}'') (P17). We extend the discussion on these limitations in \cref{sec:takeaways_and_discussion}.

\vspace{-0.5mm}
\section{Limitations and Discussion}
\label{sec:takeaways_and_discussion}
\vspace{-0.75mm}
We demonstrated that our speech disambiguation framework reduces task comprehension difficulty, increases task performance, and performs with reasonable robustness. However, our system makes a few assumptions in language, sensing, and evaluation scope. Below we outline key limitations and future work. We expand our implementation details in the Supplementary Material.

\para{Extended Evaluation Scope}
We showed how graph-based referent disambiguation can convert spoken instructions into grounded guidance and improve instruction comprehension. However, to better reflect practical remote assistance scenarios, we plan to move beyond lab-controlled, single-turn prompts and synthetically generated utterances (simple tasks), which may not fully capture the variability and pragmatics of natural communication (``\textit{Some instructions were not how I'd describe things}'' (P17)--phrases becoming overly convoluted). We plan to derive instruction patterns from realistic bi-directional assistance session transcripts, and in 3D remote scenarios to observe any behavioral shifts from users, and conduct a more systematic evaluation with more participants.

\para{Communication Pattern Coverage}
We conducted the formative study to extend the established findings 
on spatial referencing patterns to a remote assistive scenario, and categorized the recurring patterns. However, we limit the scope of these patterns to object-centered referencing. In our studies, we found that users use view-centered (``object to my left''), or environment-centered (``near the wall'') referencing strategies as well. Furthermore, our formative study accounts for only cumulative speech patterns, instead of separating initial description patterns and recovery attempts (we only qualitatively report the orderings). We plan to expand the support of the graph representation beyond object-level, with deeper speech pattern analysis and their transition strategies. Also, we will analyze the speech patterns of users, of varying technical background.

\para{Visual Guidance Design Space}
Our core novelty lies in disambiguating referents from speech. The visual guidance is a byproduct of referent resolution. However, the effectiveness of this guidance can depend on how it is represented (dimensionality: 2D, 3D; type: arrow, circle)~\cite{hepperle20192d, zhao2025guided}. Also, limited resolution density and graphics fidelity can reduce text legibility in AR. As a next step, we will systematically separate and evaluate the design factors, and explore richer visual encoding (e.g., icon, figures)~\cite{li2025satori}, for a more informed AR guidance design and improved pipeline.

\para{Importance of Multi-modality}
Practical remote assistance situations include hands-occupied tasks (e.g., holding tools, fixing parts) where additional manual interaction is undesirable, and speech-only becomes the most practical channel. This highlights why a speech-only mode is not simply a weaker interface but a necessary operational mode in certain cases. However, providing additional cues (e.g., gaze, gesture) strengthens the system's user context comprehension~\cite{hu2025gesprompt, lee2024gazepointar, lee2025walkie}, enabling it to better assist the user. In our future work, we will explore adaptive modality switching strategies~\cite{lee2025sensible} based on inferred hand availability and situational constraints.

\para{Handling Complex Spatial Referencing}
Our system showed limitations when confronted with advanced chained expressions that required multi-step reasoning such as nested or ordinal references (``\textit{in between}'', or ``\textit{second to the left of}''). In addition, our spatial relations are limited to session axis-aligned 6DoF pose. We plan to address this with multi-hop reasoning capabilities by iteratively traversing through nodes for global relational understanding. 

\para{Robustness in the Wild}
The present implementation assumes synchronized AR session coordinates and reliable sensing, conditions that are difficult to guarantee in real-world deployments. Furthermore, the spatial references can vary by the captured viewpoint. The spatial relation of object A (which is on the left of object B) can shift with the viewer's perspective. If object A is viewed from the other side, object A is to the right of object B. These constraints challenge the scaling of our system beyond laboratory conditions. We plan to address this by introducing adaptive referencing strategies, capturing the viewpoint of the user upon relational graph initialization and applying LLM-based spatial reasoning over geometric position-based spatial layout computation. 

\para{User Agency and Transparency}
Summarized guidance effectively reduced cognitive demand by distilling lengthy utterances into concise directives anchored to referents. However, condensation risks omitting semantic nuance that may be essential in complex verbal instructions (33\% of the participants preferred \fullc over \summaryc). For future work, we plan to adapt \sys to provide a heightened level-of-control on the visual feed of the transcription to the users, as well as providing heightened transparency into the reasoning process (e.g., failure cause). This will allow users to adjust the balance between clarity and comfort according to their task demands, and adjust verbal communication strategies.

\vspace{-0.5mm}
\section{Conclusion}
\vspace{-1mm}
We presented \sys, a referent disambiguation system that converts verbal instructions into spatially grounded AR guidance for remote assistance. Motivated by our formative study showing the potential for extending recurring spatial referencing patterns to remote guidance scenarios, \sys interprets how people specify targets through speech, and resolves under-specified references without relying on manual expert annotation or other cues such as gaze or gesture. We parse instructions into representative reference types and ground them in an object-centric relational graph to disambiguate the intended referent, then situate a visual indicator with a concise directive. Our evaluation showed that our pipeline makes verbal instructions easier to follow--reduced completion time, improved accuracy, lowered perceived difficulty, and workload compared to verbal-only guidance. We also demonstrated system robustness and broader applicability through case studies. We suggested that our system can be a lightweight bridge from verbal to visually explainable, actionable guidance.
\vspace{-1mm}

\section{Data Privacy and Ethics}
\vspace{-1mm}
Participants provided informed consent prior to the study, and their identity was anonymized. Both studies were conducted under the IRB compliance of Stony Brook University (\textit{1173920}).

\acknowledgments
{
This research was supported in part by NSF award IIS2529207 and ONR award N000142312124.
}

\bibliographystyle{abbrv-doi-hyperref-narrow}

\bibliography{references}
\end{document}